\documentclass[aps,pre,twocolumn,showpacs,groupedaddress]{revtex4}

\usepackage{graphicx,psfrag}

\def\BibTeX{{\rm B\kern-.05em{\sc i\kern-.025em b}\kern-.08em
    T\kern-.1667em\lower.7ex\hbox{E}\kern-.125emX}}

\begin{document}

\title{Computing the multifractal spectrum from time series: An algorithmic approach}

\author{K. P. Harikrishnan}
\email{kp_hk2002@yahoo.co.in}
\affiliation{Department of Physics, The Cochin College, Cochin-682 002, India} 
\author{R. Misra}
\email{rmisra@iucaa.ernet.in}
\affiliation{Inter University Centre for Astronomy and Astrophysics, Pune-411 007, India} 
\author{G. Ambika}
\affiliation{Indian Institute of Science Education and Research, Pune-411 021, India} 
\author{R. E. Amritkar}
\affiliation{Physical Research Laboratory, Navarangapura, Ahmedabad-380 009, India} 

\begin{abstract}
We show that the existing methods for computing the $f(\alpha)$ spectrum from a 
time series can be improved by using a new algorithmic scheme. The scheme relies 
on the basic idea that the smooth convex profile of a typical $f(\alpha)$ spectrum  can be 
fitted with an analytic function involving a set of four independant 
parameters. While the standard existing schemes \cite {gra1,chh} generally 
compute only an incomplete $f(\alpha)$ spectrum (usually the top portion), we 
show that this can be overcome by an algorithmic approach which is \emph 
{automated} to compute the $D_q$ and $f(\alpha)$ spectrum from a time series for 
any embedding dimension. The scheme is first tested with the logistic attractor 
with known $f(\alpha)$ curve and subsequently applied to higher dimensional cases.  
We also show that the scheme can be effectively adapted for analysing practcal time series 
involving noise, with examples from two widely different real world systems. Moreover, some 
preliminary results indicating that the set of four independant parameters may  be 
used as diagnostic measures is also included.
\end{abstract}

\pacs{05.45.Ac, 05.45.Tp, 05.45.Df}

\maketitle

\textbf {It is now well established that multifractal sets and objects abound in Nature. 
A characteristic feature of these objects is the self similarity since their formation is 
governed by subtle scaling laws. An important tool to analyse these sets is the 
$f(\alpha)$ spectrum which describes how the fractal dimensions of the interwoven sets 
with defnite singularity strength are distributed. In the recent issue of \emph {Chaos}, 
participating in the discussion ``Is the normal heart rate chaotic?'', many authors 
\cite {sas,fre} stress the importance of multifractality in the study of heart rate variability 
and suggest that it can provide a new observational window into the complexity mechanism 
of heart rate control. The study also highlights the need for evaluating new nonlinear 
parameters for a better physiologcal investigation and for finding new clinical applications. 
Here we present a novel automated 
scheme to compute the $f(\alpha)$ spectrum of a multifractal from its time series. 
We show that the scheme can be applied to synthetic as well as practical time series 
involving noise. It also provides us with an additional set of two independant 
parameters apart from the conventional $\alpha_{min}$ and $\alpha_{max}$ to characterise 
any general $f(\alpha)$ curve. The utility of these parameters from the point of view of 
diagnostic measures is also pursued by analysing a few class of physiological time series.} 

\section{\label{sec:level1}INTRODUCTION}
Multifractal sets and objects form the supporting structure of nonlinear 
phenomena, prime examples being strange attractors of chaotic dynamical systems 
\cite {hil,spr} regions of high vorticity in fully developed turbulence 
\cite {man1,ben,sre1} and fractal growth patterns \cite {hal1,hal2}. Multifractal 
analysis has also been applied in a variety of other fields, such as, to describe the 
morphologic and hydrologic characteristics of river basins \cite {rin,rig} and to 
analyse the velocities of solar wind plasma in the inner heliosphere \cite {mac}. In the 
case of chaotic dynamical systems, their long time evolution takes place in a subset of 
the phase space called \emph {strange attractors}, which are characterised by a spectrum of 
generalised dimensions $D_q$ \cite {hen} and the associated singularity spectrum 
$f(\alpha)$ \cite {jen1,hal3,fal}. The $f(\alpha)$ spectrum provides a mathematically 
precise and naturally intuitive description of the multifractal measure in terms of 
inter woven sets with singularity strength $\alpha$, whose fractal dimension is 
$f(\alpha)$.

For simple chaotic systems, such as one dimensional maps, $f(\alpha)$ spectrum can be determined analytically. To evaluate the $f(\alpha)$ spectrum from the time 
series, there are basically two methods. In the conventional method \cite 
{arn,gra1}, one first computes the $D_q$ spectrum from the time series and use 
the fact that the transformation from $D_q$ to $f(\alpha)$ is a Legendre 
transformation determined by the equations \cite {gra1,atm}
\begin{equation}
    \label{e.1}
    \alpha = {d \over {dq}} [(q-1)D_{q}] 
\end{equation}

\begin{equation}
    \label{e.2}
    f(\alpha) = q \alpha - (q-1)D_{q}  
\end{equation}
However, such a procedure is generally considered to be very difficult when done 
subjectively, as it involves first smoothing the $D_q$ curve and then Legendre 
transforming.  Moreover, the error bar from 
the smoothing procedure makes the estimation of $f(\alpha)$ more difficult  
and often, the complete spectrum cannot be recovered.
 
An alternative method has been proposed in the literature by Chhabra and Jensen [CJ] 
\cite {chh} for the evaluation of $f(\alpha)$ from a time series without resorting 
to the intermediate Legendre transform. In this method $f(\alpha)$ is computed 
directly from the slopes by plotting the normalised measures defined through 
probabilities as a function of logarithm of box size for different $q$ values. The 
method gives good results and is devoid of the difficulty of Legendre transforming in 
the conventional method. But here one needs to 
use an optimal covering of the measure and the subjective evaluation of the slopes 
can give rise to error bar directly in the $f(\alpha)$ curve. Both these methods  
generally compute only an incomplete $f(\alpha)$ spectrum (usually the top region) 
and the problem becomes worse for attractors of more than one dimension.

Here we propose an algorithmic approach to overcome these difficulties and compute 
the complete spectrum from the time series for any dimension. Our scheme is based 
on the idea that the typical convex profile of the $f(\alpha)$ spectrum can be 
fitted by an analytic function involving a set of parameters. This function can be 
inverted using Eqs.~(\ref{e.1}) and ~(\ref{e.2}) to get a smooth $D_q$ curve, which 
can in turn be fitted to the $D_q$ spectrum computed from the time series. By 
changing the parameters, the statistically best fit $D_q$ curve is chosen from which 
the final $f(\alpha)$ spectrum can be evaluated. It should be noted that we are not 
proposing any new method, but a new algorithmic approach to improve the existing 
methods.

Our approach has several new features. It avoids many of the sources of error in the 
conventional method such as smoothing the $D_q$ curve and using the polynomial fit to 
recover the complete $f(\alpha)$ curve. Moreover, the whole procedure is made into an 
\emph {automated} algorithmic scheme in the sense that once the time series is given, 
the scheme computes the $D_q$ and $f(\alpha)$ curves for the required embedding 
dimension without requiring any intermediate subjective analysis. Though the 
scheme is illustrated here for the conventional method, it can in principle be 
applied for the CJ method as well. The only difference is that the fit has to be 
performed directly on the $f(\alpha)$ spectrum computed from the time series 
rather than the $D_q$ spectrum.

Apart from the computation of the $f(\alpha)$ curve, another important outcome of our 
algorithmic approach is the result that any $f(\alpha)$ curve can, in general, be 
completely characterised with the help of four independant parameters including the 
conventional $\alpha_{min}$ and $\alpha_{max}$. From a practical point of view, this 
presents us with more options for representing the changes in the multifractal 
character of a system. Some preliminary results in this regard obtained from the analysis of 
a class of physiological time series is also included in the paper. 

Our paper is organised as follows: the algorithmic scheme is discussed in detail in 
Sec.II. The scheme is then tested in Sec.III using the time series from logistic 
attractor at the period doubling accumulation point where, the $f(\alpha)$ curve is 
known theoretically. It is then applied to some other standard chaotic attractors in 
higher dimensions. Sec.IV considers the application of the scheme to practical time 
series, where the effect of noise on the $f(\alpha)$ spectrum is also studied. 
Discussions and conclusions are given in Sec.V.

\begin{figure}
\includegraphics[width=0.9\columnwidth]{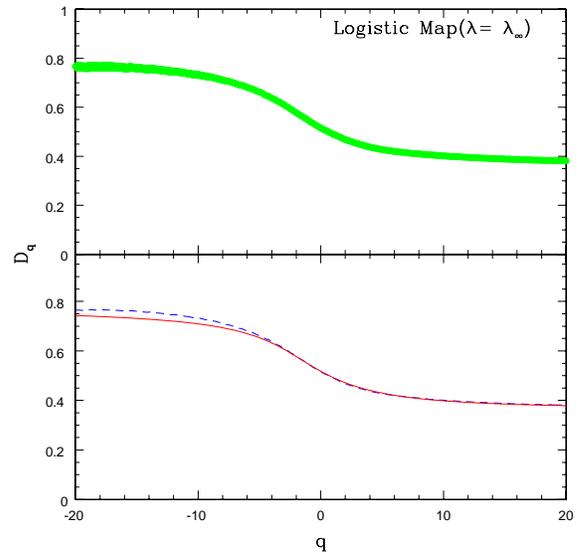}%
\caption{\label{f.1}The $D_q$ values, with error bar, of the strange attractor at the period 
doubling accumulation point of the logistic map calculated from the time series 
with $10000$ data points are 
shown in the upper panel. To show the accuracy of fitting, these values are again plotted 
in the lower panel without error bar (dashed lines) along with the 
best fit curve (continuous line).}
\label{f.1}
\end{figure}

\begin{figure}
\includegraphics[width=0.9\columnwidth]{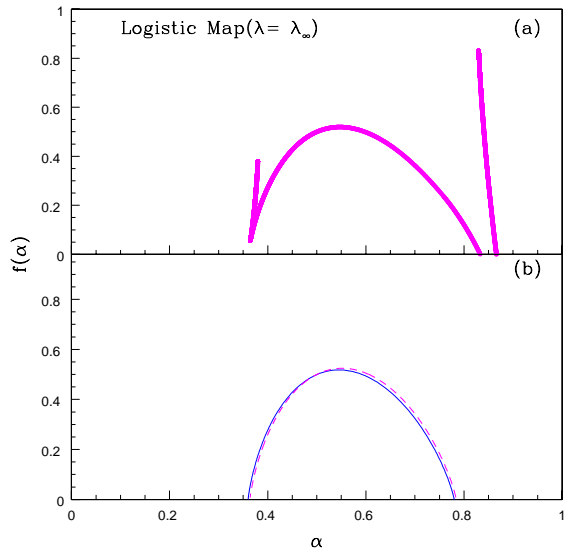}%
\caption{\label{f.2}(a)The $f(\alpha)$ spectrum of the logistic attractor obtained 
directly from the $D_q$ values which is incomplete. (b)The $f(\alpha)$ spectrum  
computed from the best fit curve (continuous line) in the previous figure 
along with the theoretical curve (dashed line). The agreement between the 
two is evident.}
\label{f.2}
\end{figure}

\section{\label{sec:level1}ALGORITHMIC SCHEME}

\subsection{Computation of $D_q$}

As the first step, the spectrum of generalised dimensions $D_q$ are 
computed from the time series using the delay embedding technique 
\cite{gra2}. For 
this, an embedded space of dimension $M$ is constructed from the scalar 
time series  $s(t_i)$ as 
\begin{equation}
   \label{e.3}
   \vec{x_i} = [s(t_i),s(t_i+\tau),....,s(t_i+(M-1)\tau)]
\end{equation}
where $\tau$ is a suitably chosen time delay. The generalised correlation sum 
$C_{q} (R)$ is  given by the relative number of data points within a distance $R$ 
from a particular ($i^{th}$) data point, say $c_{i} (R)$, raised to the power of 
$(q-1)$ and averaged over $N_c$ 
randomly selected centres:
\begin{equation}
  \label{e.4}
c_{i} (R) = {1\over N_v} \sum_{j=1,j 
\neq i}^{N_v} H(R-|\vec{x_{i}}-\vec{x_{j}}|)
\end{equation} 
\begin{equation}
   \label{e.5}
   C_{q} (R) = {1\over N_{c}} \sum_{i}^{N_{c}} c_i (R)^{q-1}
\end{equation}
where $N_v$ is the number of vectors. 
Then the spectrum of dimensions are given by
\begin{equation}
    \label{e.6}
    D_{q} \equiv   \frac {1}{q-1} \; \lim_{R \rightarrow 0} \frac{\log \; C_q (R)}{\log \; R}
\end{equation}

In practical considerations, $D_q$ is computed by taking the slope of  $ \log C_{q} (R)$ versus  $\log R$  over
a region of $R$ where the slope is nearly a constant i.e. a scaling region. For large $R$ (apart from
the formal breakdown of the limit $R \rightarrow 0$), the slope does not represent $D_q$ since the
M-spheres may extend outside  the attractor, an effect known as ``edge effect''. For small $R$, the correlation
sum is affected by counting statistics due to the finite length of the data stream. In general, the
appropriate scaling region where both these effects are negligible is often chosen by subjective visual inspection.
We have recently proposed an algorithmic scheme which non subjectively chooses an appropriate region \cite {kph}.
The scheme which was developed to compute $D_2$, uses M-cubes instead of M-spheres and chooses only those
centres where the M-cubes are inside the attractor region, thus avoiding the ``edge effect''. For large $R$,
the number of such centres decrease and by the condition that
at least $N_v/100$ centres are used for the computation, a maximum value of $R$, $R_{max}$  is obtained. 
To avoid the region dominated by counting statistics only results from $R > R_{min}$ are taken into
considerations where $N_{v} C_{2} (R) > 10$, which ensures that on the average at least ten data points 
are being considered. The above estimation of the scaling region $R_{min} < R < R_{max}$ is adequate
for computation of $D_2$ and the results obtained matches with theoretical values \cite{kph}. However,
for the generalised correlation dimensions, there is an additional error which may occur for
finite data sets. For large absolute values of  $q$, the average over the randomly chosen centres
Eq.(\ref{e.4}) may be dominated by a few centres which have either large (for $q > 0$) or small
(for $q < 0$)  values of $c_i (R)$. This biases the result towards a few centres which may be
due to statistical fluctuations. This effect is particularly strong when  $q < 0$, where centres with
statistically small values of $c_i (R)$ are the main contributors to $C_q (R)$. This is overcome by 
demanding  
that  at least $1/10$ of the centres
have a value of $c_i (R)^{q-1}$ greater than $C_{q} (R)$. This restricts the range of $R$ further and
often a suitable range of $R$ is not available for small values of $q$.  We fit a  straight line
to  $\log C_q (R)$ versus  $\log R$ for the range of $R$ that satisfy the above criteria and estimate
$D_q$ from the slope of the best fit line. The standard error on the slope is used as an estimate of
the error on $D_q$.

\subsection{Computation of $f(\alpha)$}

Attempting to compute the $f(\alpha)$ spectrum directly from the $D_q$ 
values using Eqs.~(\ref{e.1}) and ~(\ref{e.2}) leads to an incomplete $f(\alpha)$ 
spectrum (see  Fig.~\ref{f.2}a). This is mainly due to the fact that the errors in the 
calculation of $D_q$ makes the Legendre transforming numerically impractical 
because of reversal of slopes. The conventional method is to either smoothen the $D_q$ 
values or use a polynomial fit to recover the complete $f(\alpha)$ curve. Both can lead 
to large errors as has already been discussed by many authors. Here we follow a 
different procedure  as given below.

\begin{figure}
\includegraphics[width=0.9\columnwidth]{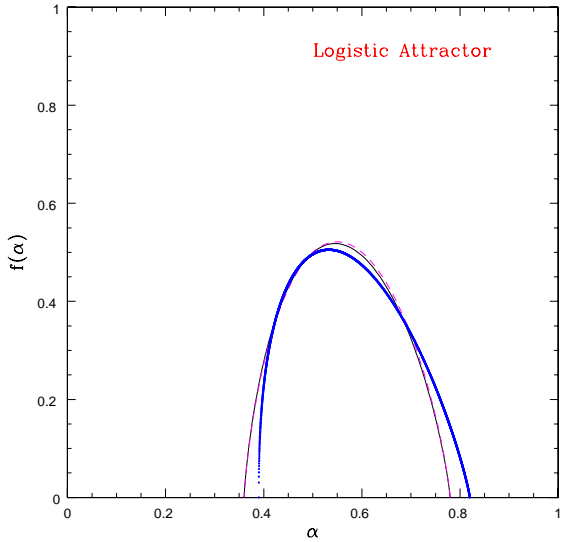}%
\caption{\label{f.3}The variation of the $f(\alpha)$ spectrum of the logistic 
attractor with number of data points. The spectrum almost completely coincide for 
$10000$ data points (dashed line) and $5000$ data points (solid line), but shows 
slight deviation as the number of data points are reduced to $3000$ (thick line).}
\label{f.3}
\end{figure}

\begin{figure}
\includegraphics[width=0.9\columnwidth]{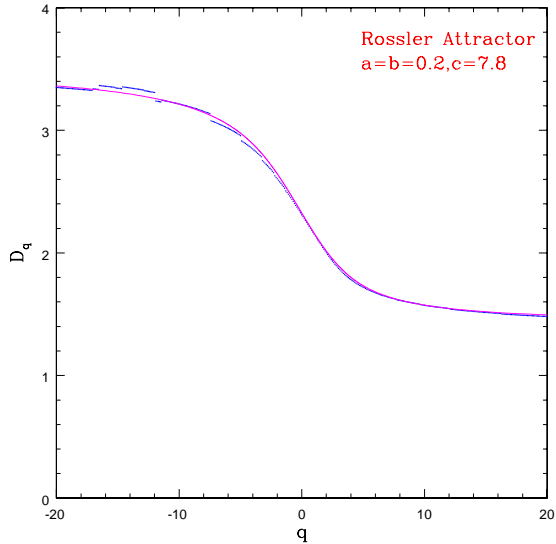}%
\caption{\label{f.4}The $D_q$ values (points) and its best fit curve (continuous line) 
for the Rossler attractor with $10000$ data points 
computed using the scheme.}
\label{f.4}
\end{figure}

\begin{figure}
\includegraphics[width=0.9\columnwidth]{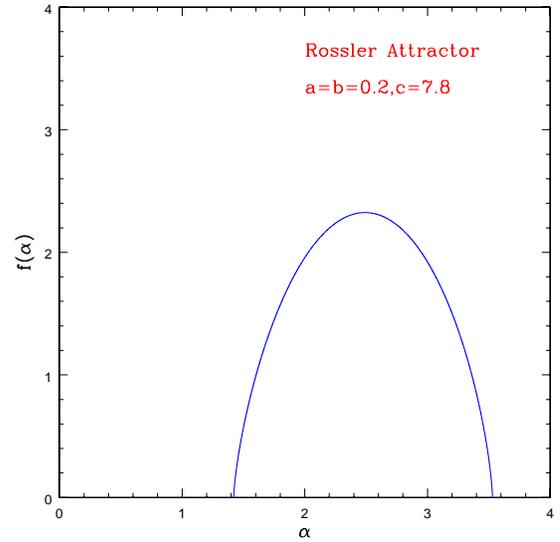}%
\caption{\label{f.5}The $f(\alpha)$ spectrum of the Rossler attractor computed from 
the best fit curve in Fig.~(\ref{f.4}).}
\label{f.5}
\end{figure}

The $f(\alpha)$ function is a single valued function defined between the
limits of $\alpha_{min}$ and $\alpha_{max}$. 
Since the derivative $f^\prime (\alpha) = d f(\alpha)/d\alpha = q$ 
is also single valued, it follows that $f (\alpha)$ has a single extremum (i.e. a maximum).
Moreover, $f(\alpha_{min}) = f (\alpha_{max}) = 0$ and $f^\prime (\alpha_{min})$ and
$f^\prime (\alpha_{max})$ tend to $ \infty$ and $ - \infty$ respectively. A simple function
which can satisfy all the above necessary conditions is 
\begin{equation}
   \label{e.7}
   f(\alpha) = A(\alpha - \alpha_{min})^{\gamma_1}(\alpha_{max} - \alpha)^{\gamma_2}
\end{equation} 
where $A$, $\gamma_1$, $\gamma_2$, $\alpha_{min}$ and $\alpha_{max}$ are a 
set of parameters characterising a particular $f(\alpha)$ curve. We will show that 
out of these five parameters, only four are independant which can unambiguously fix 
any general  $f (\alpha)$ curve. 
From  Eq.~(\ref{e.2}), we get
\begin{equation}
   \label{e.8}
   q = {d \over {d\alpha}}f(\alpha)
\end{equation}
Substituting for $f(\alpha)$ from above and simplifying
\begin{equation}
   \label{e.9}
   q = f(\alpha) \left [{{\gamma_1} \over {\alpha - \alpha_{min}}} - {{\gamma_2} \over {\alpha_{max} - \alpha}} \right ]
\end{equation}
The form of $f(\alpha)$ assumed in  Eq.~(\ref{e.7}) implies that for it to be a 
well behaved function, $A$ should be positive and $\gamma_1, \gamma_2 > 0$. 
If $\gamma_1, \gamma_2 < 0$, then $f(\alpha) \rightarrow \infty$, as $\alpha \rightarrow \alpha_{min}, \alpha_{max}$. 
Imposing the condition that the slope of $f(\alpha)$ (that is, ${d \over {d\alpha}}f(\alpha)$) should be $\infty$ at $\alpha = \alpha_{min}, \alpha_{max}$, 
we find from  Eq.~(\ref{e.9}) that this is possible only if both 
$\gamma_1, \gamma_2 < 1$. Thus the range of $\gamma_1, \gamma_2$ should be 
restricted to
\begin{equation}
   \label{e.10}
   0 < \gamma_1, \gamma_2 < 1
\end{equation}
Corresponding to $q = 1$, there is a value of $\alpha (\equiv \alpha_1)$ and 
$f(\alpha) (\equiv f(\alpha_1))$ such that 
\begin{equation}
   \label{e.11}
   D_1 = \alpha_1 = f(\alpha_1)
\end{equation}
Putting $q = 1$ in  Eq.~(\ref{e.9}), we get
\begin{equation}
   \label{e.12}
   \alpha_1 \left [{{\gamma_1} \over {\alpha_1 - \alpha_{min}}} - {{\gamma_2} \over {\alpha_{max} - \alpha_1}} \right ] = 1
\end{equation}
Using $\alpha_1, \alpha_{min}, \alpha_{max}$ and $\gamma_1$ as input 
parameters, $\gamma_2$ can be calculated from this equation. These values are 
then used to calculate the parameter $A$ from the original $f(\alpha)$ fit 
$(Eq.~(\ref{e.7}))$ with $\alpha = \alpha_1 (\equiv  f(\alpha_1))$. 
Thus only four independant parameters are required to fix the $f (\alpha)$ curve.

The scheme first takes $\alpha_1 (\equiv D_1), \alpha_{min} (\equiv D_{\infty})$ 
and $\alpha_{max} (\equiv D_{-\infty})$ as input parameters from the 
computed $D_q$ values and choosing an initial value for $\gamma_1$ in the 
range $[0,1]$, the parameters $\gamma_2$ and $A$ are calculated. The 
$f(\alpha)$ spectrum is then computed in the range $[\alpha_{min}, \alpha_{max}]$. 
From this, the $D_q$ versus $q$ curve is then computed by inverting 
 Eqs.~(\ref{e.1}) and  ~(\ref{e.2}) and fitted to the $D_q$ values computed from the 
time series.

\begin{figure}
\includegraphics[width=0.9\columnwidth]{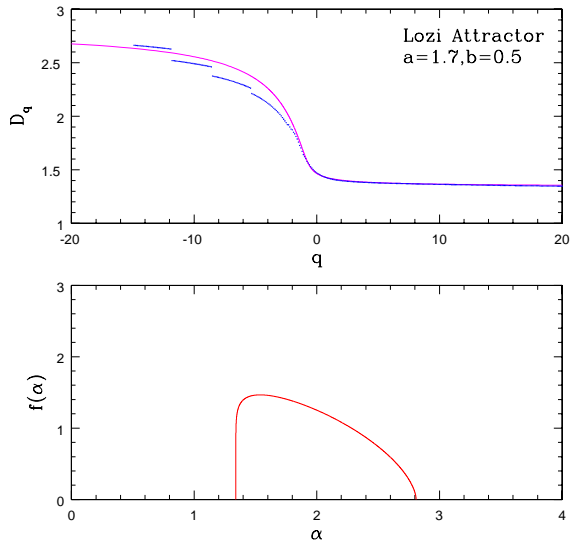}%
\caption{\label{f.6}The upper panel shows the $D_q$ values and its best fit 
curve (continuous line) for the Lozi attractor obtained by applying our 
numerical scheme. 
The corresponding $f(\alpha)$ curve is shown in the lower panel. 
Note that the $D_q$ branch for $q > 0$ is almost flat resulting in a 
highly asymmetric $f(\alpha)$ curve.}
\label{f.6}
\end{figure}

A $\chi^2$ fitting is undertaken by changing the parameter $\gamma_1$ 
(which in turn changes $\gamma_2$ and $A$) untill the functional fit 
matches with the $D_q$ values from the time series in the best possible 
manner as indicated by the minimum of the $\chi^2$ value. The final 
$f(\alpha)$ spectrum is computed from the functional fit.

In the scheme, $D_1$ is used as one of the input parameters as it is obtained 
directly from the $D_q$ spectrum computed from the time series. But to represent the 
changes in the $f (\alpha)$ curve, it is more convenient to use $\gamma_1$ and 
$\gamma_2$ along with $\alpha_{min}$ and $\alpha_{max}$ as the free parameters. 
While $\alpha_{min}$ and $\alpha_{max}$ fix the end points of the spectrum (at which 
the slope becomes $\infty$), the values of $\gamma_1$ and $\gamma_2$ completely 
specify the nature of the $f(\alpha)$ profile. For example, as the difference 
between $\gamma_1$ and $\gamma_2$ increases, the spectrum becomes more and more 
asymmetric between the two branches. Moreover, they also determine the peak value of 
the spectrum $D_0$. If $\gamma_1 \sim \gamma_2 \sim 0$, then $D_0 \ll M$ and on the 
other hand, if $\gamma_1 \sim \gamma_2 \sim 1$, then $D_0 \sim M$. Thus the four 
parameters $\alpha_{min}, \alpha_{max}, \gamma_1$ and $\gamma_2$ can, in principle, 
completely characterise any  $f(\alpha)$ spectrum.

\section{\label{sec:level1}APPLICATION TO SYNTHETIC DATA}

To test our scheme, it is first applied on the time series from the logistic attractor 
at the period doubling accumulation point, where the $f(\alpha)$ spectrum is 
known theoretically. The analysis is done with $10000$ data points. The $D_q$ spectrum 
is first computed using Eq.~(\ref{e.6}) (with 
$M=1$), for $q$ values in the range $[-20,+20]$. The computation is done taking a 
step width of $\Delta q = 0.1$. Choosing $D_{-20}, D_1$ and $D_{20}$ as the input values 
for the $f(\alpha)$ function Eq.~\ref{e.7}, the parameters $\gamma_1$ and $\gamma_2$ 
are scanned in the range $[0,1]$ untill the functional fit matches the $D_q$ values 
as indicated by the $\chi^2$ minimum. Since the error in $D_q$ generally bulges as 
$q \rightarrow -20$, the error bar is also taken into account in the fitting 
process. The $D_q$ values with error bar and the best fit curve are shown in 
Fig.~\ref{f.1}. The complete $f(\alpha)$ spectrum computed from the best fit 
$D_q$ curve is shown in Fig.~\ref{f.2}. For one dimensional maps, the $f(\alpha)$ 
curve can be determined theoretically \cite {hil}. To make a comparison, the 
theoretical $f(\alpha)$ curve is superimposed on the computed one in Fig.~\ref{f.2}. 
Also shown in Fig.~\ref{f.2}a is the incomplete $f(\alpha)$ spectrum computed 
directly from the $D_q$ values.

Since the $f(\alpha)$ spectrum for the logistic attractor is exactly known, it can also 
be used to test our scheme with respect to the number of data points required in a 
time series for a reasonable estimate of the $f(\alpha)$ spectrum. This is shown in 
Fig.~\ref{f.3}, where the spectrum for the logistic attractor for three different 
number of data streams, namely, $3000$, $5000$ and $10000$ are shown. It turns out 
that, a reasonable approximation to the $f(\alpha)$ spectrum can be obtained atleast with 
$3000$ data points using our scheme in one dimension.

As the second example, we use time series generated from another standard chaotic attractor, 
namely, the Rossler attractor, for parameter values $a=0.2,b=0.2,c=7.8$, with a time step of 
$\Delta t=0.1$. The $D_q$ and $f(\alpha)$ spectrum are computed as above taking the 
total number of $10000$ data points. The $D_q$ values and best fit curves for 
are shown in Fig.~\ref{f.4}, while the $f(\alpha)$ curve is 
shown in Fig.~\ref{f.5}. 
 
\begin{figure}
\includegraphics[width=0.9\columnwidth]{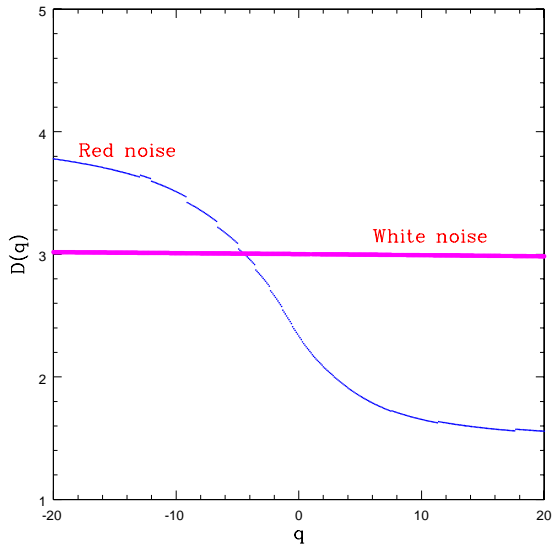}%
\caption{\label{f.7}The $D_q$ spectrum for white noise and red noise from $10000$ data 
points. Note that the latter behaves like a chaotic system with a well defined 
$D_q$ curve.}
\label{f.7}
\end{figure}

\begin{figure}
\includegraphics[width=0.9\columnwidth]{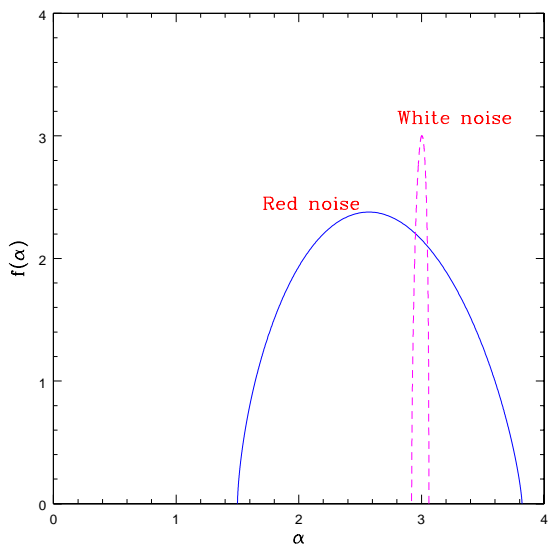}%
\caption{\label{f.8}The $f(\alpha)$ spectrum for white noise and red noise. The scheme 
accurately determines the $f(\alpha)$ spectrum of white noise which is expected to be a 
$\delta$ function corresponding to the embedding dimension.}
\label{f.8}
\end{figure}

\begin{figure}
\includegraphics[width=0.9\columnwidth]{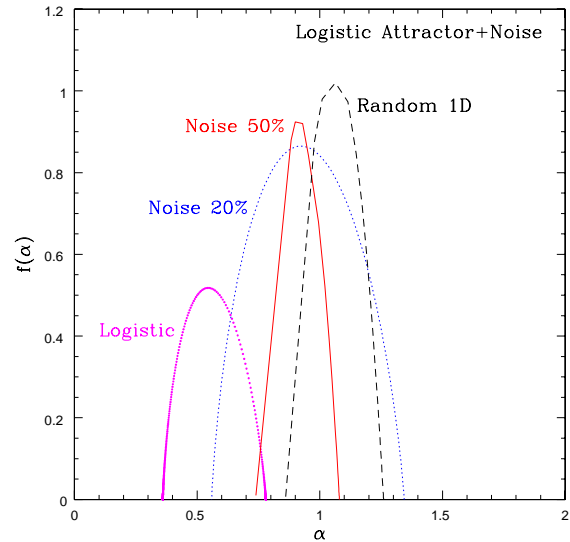}%
\caption{\label{f.9}Result of addition of white noise to the data from the 
logistic attractor. The $f(\alpha)$ spectrum for a noise level of $20\%$ (dotted 
line) and $50\%$ (solid line) are shown along with that of logistic attractor 
(points). The number of data points used for computation are $10000$. Note that 
for $50\%$ of noise contamination, the peak of the spectrum $(D_0)$ almost 
touches the embedding dimension 1. For comparison, the $f(\alpha)$ spectrum 
of pure random noise in one dimension is also shown (dashed line).}
\label{f.9}
\end{figure}

In the above two cases, the $f(\alpha)$ curves are almost symmetric between the two 
branches. We now show that our scheme is also useful for the computation of more 
general types of $f(\alpha)$ curves. An example is that of the standard Lozi 
attractor. It is found that the $D_q$ spectrum in this case is almost flat for 
$q > 0$. But the $D_q$ spectrum can be accurately fitted by using the standard form of 
the $f(\alpha)$ function, as shown in Fig.~\ref{f.6}. The corresponding $f(\alpha)$ 
spectrum computed from the best fit $D_q$ curve is also shown in Fig.~\ref{f.6} 
(lower panel), which turns out to be highly asymmetric between the two branches.
While the parameter values $\gamma_1$ and $\gamma_2$ for the Rossler attractor 
are very close $(\gamma_1 = 0.65, \gamma_2 = 0.60)$ as reflected in the nearly 
symmetric  $f(\alpha)$ profile, that for Lozi attractor are completely different 
with $\gamma_1 = 0.09$ and $\gamma_2 = 0.60$. The scheme has also been applied to 
compute the spectrum of other standard chaotic attractors, such as, Henon and Lorenz.

\section{\label{sec:level1}APPLICATION TO REAL WORLD DATA}

Before the scheme is applied to practical time series, it is important to test it with 
time series involving noise. Since pure white noise is scale free, one expects the 
corresponding $f(\alpha)$ spectrum to be ideally a $\delta$ function with 
$f(\alpha) \equiv \alpha = M$, the embedding dimension. In Fig.~\ref{f.7}, we show the 
$D_q$ spectrum of pure white noise and a colored noise with spectral index 2.0 
(called \emph {red noise}), computed from the respective time series for embedding 
dimension $M=3$. The $f(\alpha)$ spectrum computed for both are shown in Fig.~\ref{f.8}. 
While the $f(\alpha)$ spectrum of white noise is a $\delta$ function as expected, that of 
colored noise looks like a normal $f(\alpha)$ spectrum.
 
\begin{figure}
\includegraphics[width=0.9\columnwidth]{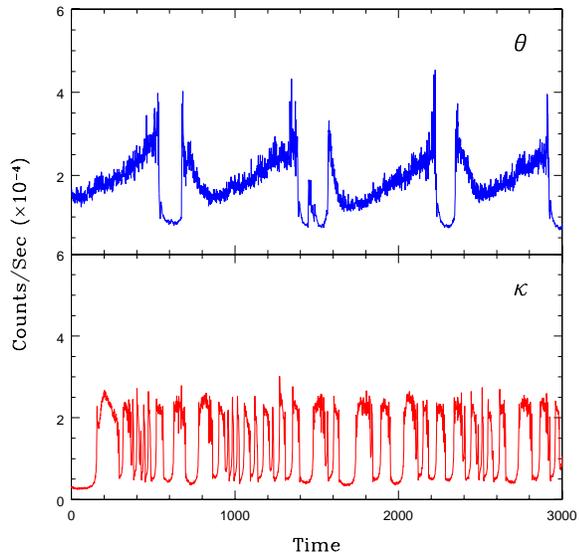}%
\caption{\label{f.10}A part of the light curves from two temporal states of the 
black hole system GRS1915+105.}
\label{f.10}
\end{figure}

To study the effect of noise on the $f(\alpha)$ spectrum of a chaotic attractor, we 
generate two time series by adding $20\%$ and $50\%$ white noise to the data from 
the logistic attractor using $10000$ data points. White noise is used since it is 
scale free and hence can significantly alter the $f(\alpha)$ spectrum, while 
colored noise behaves much like a chaotic attractor with a well defined spectrum. 
Fig.~\ref{f.9} shows the $f(\alpha)$ spectrum of the logistic attractor added with 
$20\%$ and $50\%$ white noise, along with that of the logistic attractor. As the 
percentage of noise increases, the spectrum tends more and more towards a delta 
functon, centered around the embedding dimension $M=1$. To get a proper 
comparison, the figure also shows the $f(\alpha)$ spectrum of pure random noise 
in one dimension.

\begin{figure}
\includegraphics[width=0.9\columnwidth]{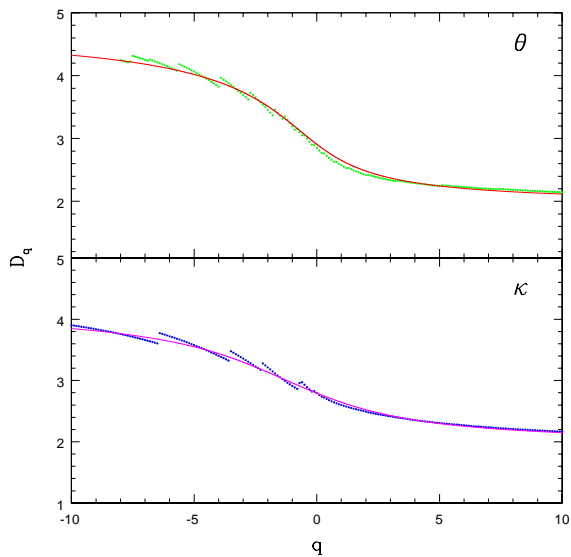}%
\caption{\label{f.11}The $D_q$ spectrum along with the best fit curve for the two states 
$\theta$ and $\kappa$ of the black hole system for embedding dimension $M = 3$.}
\label{f.11}
\end{figure}

\begin{figure}
\includegraphics[width=0.9\columnwidth]{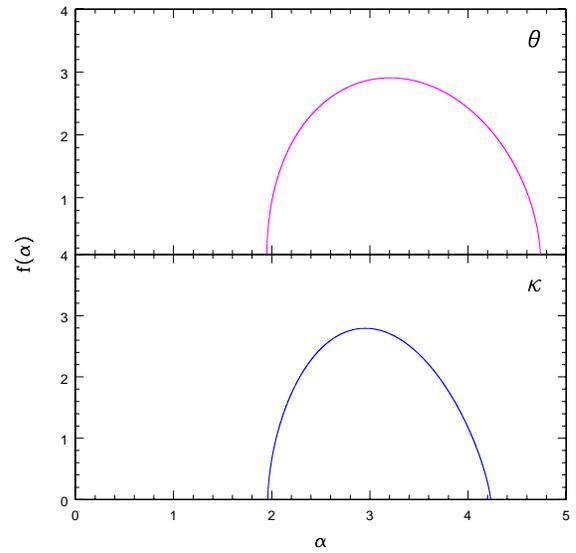}%
\caption{\label{f.12}The $f(\alpha)$ spectrum for the black hole states computed from 
the best fit $D_q$ curves in the previous figure.}
\label{f.12}
\end{figure}

We now apply our scheme to some real world data. We choose time series from two 
important fields, namely, astrophysics and physiology, where methods and concepts from 
nonlinear dynamics are constantly being applied. The first example is the X-ray light curve 
from a prominent back hole binary, GRS1915+105. The light curves from this black hole 
system have been classified into 12 temporal states by Belloni et.al \cite {bel} 
based on RXTE observation. Here we choose data from two representative classes, 
$\theta$ and $\kappa$, and generate continuous light curves of approximately $6000$ 
data points for both class. Fig.~\ref{f.10} shows a part of the light curves used for 
the analysis. Using surrogate analysis, we have recently shown \cite {mis1,mis2} that 
light curves from more than half of the 12 temporal states (including $\theta$ and 
$\kappa$) show significant deviation from stochastic behavior. The saturated value of 
correlation dimension for both are $<3$. Hence $M=3$ is chosen for applying our 
numerical scheme. The result of applying our scheme to the two light curves are shown 
in  Fig.~\ref{f.11} and   Fig.~\ref{f.12}. The former shows the computed $D_q$ 
spectrum along with the best fit curve, while the latter shows the $f(\alpha)$ 
spectrum computed from the best fit curves. Note that, though the $D_q$ values are 
discontinuous for large negative $q$ values, one can statistically fit a smooth 
curve for $D_q$ from which the $f(\alpha)$ spectrum can be derived. 

\begin{figure}
\includegraphics[width=0.9\columnwidth]{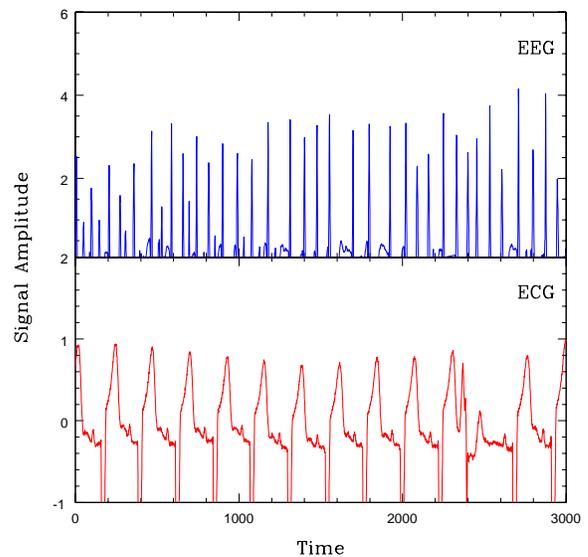}%
\caption{\label{f.13}A part of the EEG and ECG signals analysed in this work.}
\label{f.13}
\end{figure}

\begin{figure}
\includegraphics[width=0.9\columnwidth]{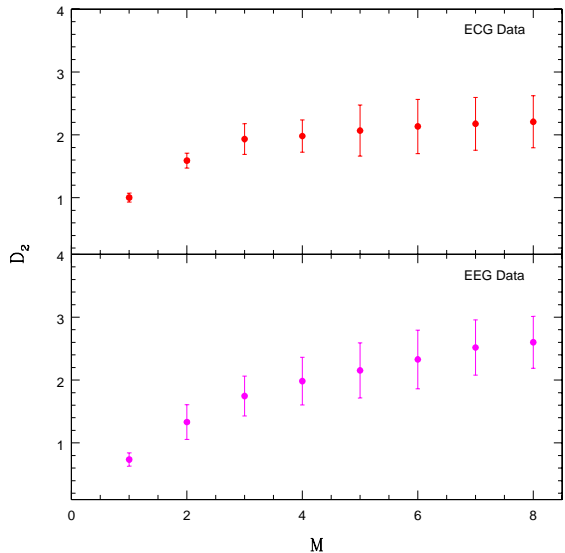}%
\caption{\label{f.14}The variation of correlation dimension $D_2$ (with error bar) 
as a function of $M$ for the EEG and ECG signals.}
\label{f.14}
\end{figure}

As the second example, we use two data sets from physiology, namely an EEG data and 
an ECG data. The EEG data was downloaded from the website of the Department of 
Epileptology, University of Bonn while the ECG data was obtained from 
http://www.physionet.org/physiobank/archives. The EEG data is from an epileptic 
patient during seizure activity. The data consists of continuous data streams of 
about 24 seconds long and consisting of approximately $5000$ data points. The 
ECG data was recorded from a heart patient with a congestive disorder and consists of 
continuous data streams of $5400$ data points with a sampling time of $0.04$ seconds. 
Both signals are shown in  Fig.~\ref{f.13}. First we compute the correlation 
dimension $D_2$ of both signals by applying the nonsubjective scheme \cite {kph} 
recently proposed by us, with the result shown in  Fig.~\ref{f.14}. For both signals, 
$D_2$ saturate well below $M = 3$. The results of applying our $f(\alpha)$ scheme 
are shown in  Fig.~\ref{f.15} and  Fig.~\ref{f.16}, with multifractal character 
evident in both cases. Thus it is clear that the 
scheme can be successfully employed to compute the $D_q$ and $f(\alpha)$ spectrum 
from practical time series of finite data streams even with noise contamination, 
provided there exists an underlying chaotic attractor.

\begin{table*}
\caption{\label{t.1}The parameter values computed by our scheme for physiological data sets 
corresponding to four different class. The average values of five data streams for 
each class are shown.}
\begin{ruledtabular}
\begin{tabular}{cccccc}
\hline
\emph{Data Class}  &  $\alpha_{min}$ & $\alpha_{max}$  &  $\gamma_1$  &  $\gamma_2$  &  $|\gamma_1 - \gamma_2|$  \\
\hline

EEG               &                 &              &              &                &            \\
Healthy        & $1.71 \pm 0.08$ & $4.18 \pm 0.17$  & $0.37 \pm 0.10$ & $0.32 \pm 0.14$ & $0.05$     \\  

& & \\

EEG               &                 &              &              &                &            \\
Epileptic Seizure  & $1.28 \pm 0.06$ & $2.85 \pm 0.14$  & $0.34 \pm 0.08$ & $0.05 \pm 0.03$ & $0.29$     \\  

& & \\

ECG               &                &              &              &                &            \\
Healthy        & $1.46 \pm 0.12$ & $4.30 \pm 0.22$  & $0.69 \pm 0.08$ & $0.58 \pm 0.05$ & $0.11$         \\  

& & \\

ECG               &               &              &              &                 &           \\
Congestive Heart Failure & $1.88 \pm 0.10$ & $4.07 \pm 0.18$ & $0.29 \pm 0.06$ & $0.08 \pm 0.04$ & $0.21$   \\  
\hline
\end{tabular}
\end{ruledtabular}
\end{table*}

\begin{figure}
\includegraphics[width=0.9\columnwidth]{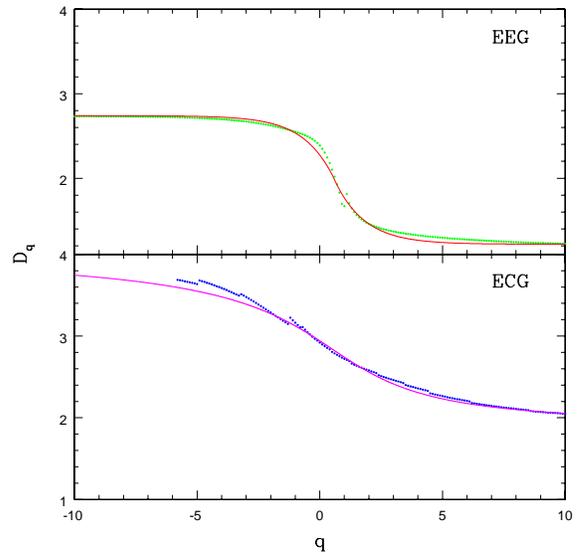}%
\caption{\label{f.15}The $D_q$ spectrum and the best fit curves for the EEG and ECG signals 
for $M = 3$.}
\label{f.15}
\end{figure}

\begin{figure}
\includegraphics[width=0.9\columnwidth]{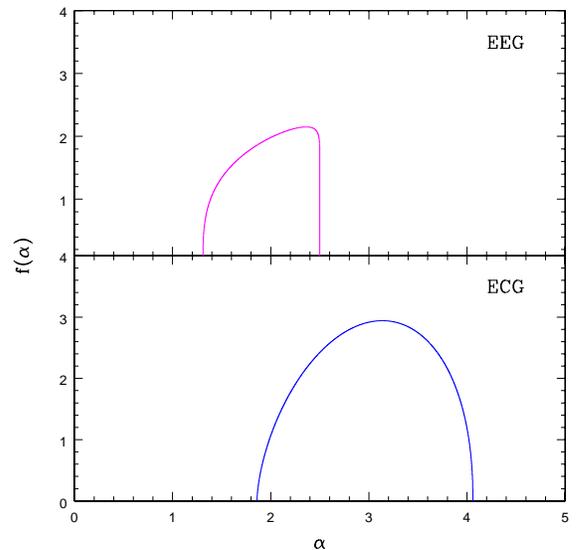}%
\caption{\label{f.16}The $f(\alpha)$ spectrum for the EEG and ECG signals computed from the 
best fit $D_q$ curves in the previous figure.}
\label{f.16}
\end{figure}

Finally, we would also like to stress the importance of computing the $f(\alpha)$ spectrum 
using an automated scheme, such as the one presented here. In the four examples of real world data 
that we have analysed above, the saturated $D_2$ values are approximately same, but the 
$f(\alpha)$ spectra are quite different. The subtle changes in the $f(\alpha)$ spectra can be 
better studied using the present scheme. This is because, the scheme provides an additional 
set of two independant parameters $\gamma_1$ and $\gamma_2$, apart from $\alpha_{min}$ and  
$\alpha_{max}$  which typically characterise the changes in the $f(\alpha)$ profile. The 
utility of these parameters can be seen from Table ~\ref{t.1}, which shows the result of a 
preliminary study made on some limited physiological time series. We have analysed four class 
of physiological data sets downloaded from the above mentioned websites. They are EEG and ECG 
signals from healthy human beings, EEG signal during epileptic seizure and ECG signals from 
patients with congestive heart failure. We have analysed five data streams for each class and 
Table ~\ref{t.1} shows the average values of the two parameters $\gamma_1$ and $\gamma_2$ 
for each class along with $\alpha_{min}$ and $\alpha_{max}$, with error 
bar showing the range of variation. We find that the values of $|\gamma_1 - \gamma_2|$ 
are significantly different for the healthy data and those with physiological disorder in 
both cases. Thus, just like $|\alpha_{max} - \alpha_{min}|$ which is conventionally used to 
characterise the geometric complexity of a multifractal, the quantity $|\gamma_1 - \gamma_2|$ 
may also be a potential candidate to quantify the inherent changes in the multifractal 
character of the system. 
Ofcourse, these results are only preliminary and have to be confirmed using a 
much larger number of data sets and requires an extensive analysis. Nevertheless, the initial 
results indicate that our scheme can be efficiently employed to quantify the changes in the 
multifractal character between various states of a complex system (such as the black hole system) 
or track changes in the $f(\alpha)$ spectrum arising out of various physiological 
disorder as reflected in the ECG or EEG time series, as shown here. We hope the scheme 
can be better utilised in this regard.
 
\section{\label{sec:level1}DISCUSSION AND CONCLUSION}

Computing the multifractal spectrum of a chaotic attractor from its time series is 
generally considered to be a difficult task. Often, only a part of the $f(\alpha)$ 
spectrum can be recovered numerically from the $D_q$ curve due to various reasons 
such as, the errors in the computation of $D_q$ and the Legendre transforming involved. 
Here we show that the existing methods can be improved  using a new algorithmic 
approach by which the complete $f(\alpha)$ spectrum can be evaluated from a 
time series.

The scheme first assumes an analytical function for the $f(\alpha)$ curve, from which a 
functional fit for the computed $D_q$ values can be obtained by the inverse Legendre 
transform. The best fit curve is then used to derive the complete $f(\alpha)$ spectrum. 
The scheme is illustrated using time series from standard low dimensional chaotic 
systems and then applied to a variety of practical data from real world.  

We have recently shown \cite{kph2} that the the $f(\alpha)$ gives information only 
upto two scales evenif the underlying multiplicative process involves more than two. 
As a consequence, the $f(\alpha)$ spectrum of a chaotic attractor can, in general, be 
mapped onto that of a two scale Cantor set. This also provides an alternative method 
for computing the $f(\alpha)$ spectrum and to characterise a chaotic attractor 
in terms of the three independant parameters of a two scale Cantor set.

In contrast, an important aspect of the present  
scheme is that, an analytic function is 
proposed (which is probably unique) to fit all convex $f(\alpha)$ curves in general.
Since the whole process is automated and the analysis is done under identical conditions 
prescribed by the algorithmic scheme, the resulting parameters characterising the 
spectrum (derived from a fitting function for $f(\alpha)$) give a better representation 
for comparison between data sets. This is especially 
important in the case of real world data since the changes in the same system, such as  
for example, due to some changes in the parameter, can be compared in a 
non-subjective manner.

In order to show this explicitely, the analysis of a few class of physiological data is 
also presented. We find that the subtle variations in the $f(\alpha)$ curve can be better 
characterised using our algorithmic scheme. Though our results indicate that the parameters 
may also be useful from a diagnostic point of view, this requires a much more 
comprehensive analysis using large number of data sets for confirmation. This is 
currently under way and will be presented elsewhere.

\begin{acknowledgments}
The authors thank the Department of Epileptology, University of Bonn, for making the human 
brain EEG data available on their website.

KPH and RM acknowledge the financial support from Dept. of Sci. and Tech., Govt. of India, 
through a Research Grant No. SR/S2/HEP - 11/2008.

KPH  acknowledges the hospitality and computing facilities in IUCAA, Pune.
\end{acknowledgments}

\end{document}